\documentclass[onecolumn,trackchanges]{aastex631}

\newcommand{\chn}{{\it Chandra}}

\shorttitle{3CR\,403.1 ApJ}
\shortauthors{Missaglia et al.}
\graphicspath{{./}{}}

\begin{document}
\nocite{*}
	
	\title{High frequency radio imaging of 3CR 403.1 with the Sardinia Radio Telescope}
	
	\correspondingauthor{Valentina Missaglia}
	\email{valentina.missaglia@unito.it}
	
	\author[0000-0001-8382-3229]{Valentina Missaglia}
	\affiliation{Dipartimento di Fisica, Universit\`a degli Studi di Torino, via Pietro Giuria 1, I-10125 Torino, Italy.}
	\affiliation{INAF-Osservatorio Astrofisico di Torino, via Osservatorio 20, I-10025 Pino Torinese, Italy.}
	\affiliation{INFN-Istituto Nazionale di Fisica Nucleare, Sezione di Torino, I-10125 Torino, Italy.}
	
	\author[0000-0002-4800-0806]{Matteo Murgia}
	\affiliation{INAF-Osservatorio Astronomico di Cagliari, Via della Scienza 5 - I-09047 Selargius (CA), Italy}
	
	\author[0000-0002-1704-9850]{Francesco Massaro}
	\affiliation{Dipartimento di Fisica, Universit\`a degli Studi di Torino, via Pietro Giuria 1, I-10125 Torino, Italy.}
	\affiliation{INAF-Osservatorio Astrofisico di Torino, via Osservatorio 20, I-10025 Pino Torinese, Italy.}
	\affiliation{INFN-Istituto Nazionale di Fisica Nucleare, Sezione di Torino, I-10125 Torino, Italy.}
	
	\author[0000-0002-5646-2410]{Alessandro Paggi}
	\affiliation{Dipartimento di Fisica, Universit\`a degli Studi di Torino, via Pietro Giuria 1, I-10125 Torino, Italy.}
	\affiliation{INAF-Osservatorio Astrofisico di Torino, via Osservatorio 20, I-10025 Pino Torinese, Italy.}
	\affiliation{INFN-Istituto Nazionale di Fisica Nucleare, Sezione di Torino, I-10125 Torino, Italy.}
	
	\author[0000-0003-4413-7722]{Ana Jimenez-Gallardo}
	\affiliation{Dipartimento di Fisica, Universit\`a degli Studi di Torino, via Pietro Giuria 1, I-10125 Torino, Italy.}
	\affiliation{INAF-Osservatorio Astrofisico di Torino, via Osservatorio 20, I-10025 Pino Torinese, Italy.}
	\affiliation{INFN-Istituto Nazionale di Fisica Nucleare, Sezione di Torino, I-10125 Torino, Italy.}
	
	\author[0000-0002-9478-1682]{William R. Forman}
	\affiliation{Center for Astrophysics $\mid$ Harvard \& Smithsonian, 60 Garden Street, 02138, Cambridge (MA), USA}
	
	\author[0000-0002-0765-0511]{Ralph P. Kraft}
	\affiliation{Center for Astrophysics $\mid$ Harvard \& Smithsonian, 60 Garden Street, 02138, Cambridge (MA), USA}
	
	\author[0000-0002-0690-0638]{Barbara Balmaverde}
	\affiliation{INAF-Osservatorio Astrofisico di Torino, via Osservatorio 20, I-10025 Pino Torinese, Italy.}

	\begin{abstract}
	We present multifrequency observations of the radio source 3CR\,403.1, a nearby (z=0.055), extended ($\sim$0.5 Mpc) radio galaxy hosted in a small galaxy group. Using new high frequency radio observations from the Sardinia Radio Telescope (SRT), augmented with archival low frequency radio observations, we investigated radio spectral and polarimetric properties of 3CR\,403.1. From the MHz-to-GHz spectral analysis, we computed the equipartition magnetic field in the lobes to be B$_{eq}$=2.4~$\mu$G and the age of the source to be $\sim$100 Myr. From the spectral analysis of the diffuse X-ray emission we measured the temperature and density of the intracluster medium (ICM). From the SRT observations, we discovered two regions where the radio flux density is below the background value. We computed the Comptonization parameter both from the radio and from the X-ray observations to test if the Sunyaev-Zel'dovich effect is occurring here and found a significant tension between the two estimates. If the negative signal is considered as real, then we speculate that the discrepancy between the two values could be partially caused by the presence of a non-thermal bath of mildly relativistic ghost electrons. From the polarimetric radio images, we find a net asymmetry of the Faraday rotation between the two prominent extended structures of 3CR\,403.1, and constrain the magnetic field strength in the ICM to be 1.8-3.5 $\mu$G. The position of 3CR\,403.1 in the magnetic field-gas density plane is consistent with the trend reported in the literature between central magnetic field and central gas density.
	\end{abstract}
	

	\section{Introduction} 
	\label{sec:intro}
	
	The Third Cambridge Catalog (3C) of Radio Sources is a northern hemisphere sample of radio galaxies and quasars, originally detected at 159\,MHz. Its first edition was published in 1959 \citep{edge59}, while two revised versions were released in 1962 \citep[i.e., the 3CR,][]{bennett62} and later in 1983 \citep[i.e., the 3CRR,][]{laing93}, respectively, both performed at 178 MHz with the same threshold of 9 Jy for the limiting flux density. The 3C catalog, and all its revised versions \citep[see e.g.][]{1985PASP...97..932S}, are still a paramount tool to study radio loud active galactic nuclei (AGNs) and the interaction with their environments at all scales \citep[i.e., feedback processes:][]{2005xrrc.procE7.09K,2012ARA&A..50..455F,2014ARA&A..52..589H,2016MNRAS.458..681D}. 
	
	Since the early sixties, follow up multifrequency observations began, aimed at obtaining a full overview of the 3C radio sources  \citep{1995MNRAS.274..939L,1996ApJS..107..621D,1997ApJS..112..415M}. Initial observational campaigns carried out optical spectroscopic observations aimed at estimating the source redshifts \citep{1960ApJ...132..908M,1965ApJ...142.1667L,1966ApJ...145....1S,1967ApJ...150L.145S,1976PASP...88..621S,1980PASP...92..553S}. These then evolved into more complete studies with wider broadband coverage using several telescopes \citep[e.g., Jansky Very Large Array, Spitzer, Herschel, Hubble Space Telescope][]{1998ApJS..119...25H,2000A&A...355..873C,2007ApJ...660..117C,2015A&A...575A..80P,2009ApJS..183..278T,2006ApJS..164..307M}. These campaigns were recently augmented thanks to X-ray observations of the 3CR \chn\ Snapshot Survey started in 2008 \citep{2010ApJ...714..589M,2013ApJS..206....7M}. The main aim of this high energy survey is to detect X-ray emission arising from nuclei, lobes, jet knots and hotspots, as well as that of the intergalactic medium for those radio sources harbored in galaxy-rich large scale environments. \citep{2018ApJS..238...31M,2018ApJ...867...35R,2021A&A...647A..79P,2021ApJ...912L..25J}. Moreover, one of the underlying objectives of this X-ray snapshot survey is to identify new 3C sources that could merit additional follow up observations as is the case presented here: 3CR\,403.1 \citep[see e.g.,][for details on other follow up Chandra observations of 3CR sources]{2010MNRAS.401.2697H,2012MNRAS.424.1774H,2012MNRAS.419.2338O,2016MNRAS.458..681D}.
	
	During a follow up analysis of several targets of the 3CR \chn\ Snapshot Survey, we discovered extended X-ray emission around this nearby radio galaxy \citep{2012ApJS..203...31M,2021ApJS..252...31J}, at larger scales with respect to previous literature analyses. 
	
	3CR\,403.1 (a.k.a. 4C\,-01.51) is a classical FR\,II radio galaxy \citep{1974MNRAS.167P..31F} at $z=0.055$. As reported in \citet{1966AuJPh..19..559B} the optical counterpart is elliptical, and it has a low-ionization galaxy-like optical spectrum \citep{2010ApJ...725.2426B}. H$\alpha$ and [OIII]$\lambda$5007 luminosities are in the range between $10^{39}$ and $10^{40}$ erg s$^{-1}$ \citep{2011A&A...525A..28B}, while the radio luminosity at 178 MHz is log$_{10}(\text{P}_{178})=32.98$ \citep{1985PASP...97..932S}. In radio archival images, 3CR\,403.1 shows two prominent radio lobes, extending in the north-west to south-east direction on a scale of $\sim$ 0.5 Mpc and thus being the second most extended 3C source among the 37 sources at redshift $z<0.1$. In particular, the large scale structure of 3CR\,403.1, including the two lobes, is visible at 1.4 GHz in the NRAO VLA Sky Survey \citep[NVSS,][]{1998AJ....115.1693C} and at lower frequencies at 74 MHz in the Very Large Array Sky Survey Redux \citep[VLSSr,][]{2014MNRAS.440..327L} observations, as well as in the GaLactic and Extragalactic All-sky MWA (GLEAM) Survey, which is a continuum survey conducted using the Murchison Widefield Array between 72 and 231 MHz \citep[][]{2015PASA...32...25W}. At all radio frequencies, the radio core and lobes are detected above 5$\sigma$ level of confidence. The radio morphology of 3CR\,403.1 is peculiar since at low radio frequencies (i.e., archival VLA P band observation, 230-470 MHz) it shows two radio knots elongated on the west-east direction, separated by 105 kpc, and perpendicular to the radio axis marked by its large scale structure detected at low frequencies, that correspond to small scales cavities in the X-rays \citep{2021ApJS..252...31J}. This emission lies also at the opposite sides of the radio core, detected in the Very Large Array Sky Survey at 2-4 GHz \citep[VLASS,][]{lacy2020}. In FR\,II sources, as shown in \citet{2011ApJS..197...24M}, X-ray emission is usually detected along the large scale radio structure. Such emission is typically thought to be due to X-rays produced by relativistic electrons in the lobes that up-scatter ambient cosmic microwave background (CMB) photons via inverse-Compton scattering (IC/CMB) from lobes \citep[see e.g.][]{2002ApJ...565..244H}. However, diffuse X-ray emission surrounding 3CR 403.1 is detected perpendicular to the large scale radio structure and, thus, it is most likely due to thermal emission from the intracluster medium (ICM) \citep[see e.g. results in][]{2021ApJS..252...31J}. Therefore, given its proximity, 3CR\,403.1 is a good candidate to study the relationship between the ambient ICM and the radio structure at comparatively high spatial resolution.

	Here we present follow up observations of 3CR\,403.1 with the Sardinia Radio Telescope \citep[SRT;][]{bolli2015,prandoni2017}. These observations were carried out at the end of 2019 to (1) investigate the spectral shape at higher frequency of the radio structure, (2) perform a spatially resolved, spectral analysis at radio frequencies, given the clear detection of the radio components shown in Figure~\ref{radio_combo}, and (3) carry out radio polarimetric observations. Polarimetric radio observations, in combination with X-ray observations, permit the study of the properties of the magnetic field permeating the ICM in which radio lobes are embedded. 	
	
	This work is organized as follows. In \S~\ref{sec:reduc} we provide details about the available archival radio data and the data reduction procedure adopted for SRT and \chn\ observations. In \S~\ref{sec:results} we report a brief optical overview of the source and we discuss (i) the radio spectral analysis, (ii) the test performed to confirm the presence of a possible Sunyaev-Zel'dovich (SZ) effect, (iii) the measurements of the magnetic field by means of the rotation measure (RM). Finally, \S~\ref{sec:conclusion} is dedicated to our summary, conclusions and future perspectives.

	We adopt cgs units for numerical results and we assume a flat cosmology with $H_0=69.6$ km s$^{-1}$ Mpc$^{-1}$, $\Omega_\mathrm{M}=0.286$ and $\Omega_\mathrm{\Lambda}=0.714$ \citep{2014ApJ...794..135B}, through the whole manuscript and unless otherwise stated. Thus, according to these cosmological parameters, 1\arcsec\ corresponds to 1.076 kpc at the 3CR\,403.1 redshift (i.e., $z_\mathrm{src}=$0.055), while its luminosity distance is 246.9 Mpc. Spectral indices $\alpha$ are defined by flux density $S_\nu \propto 
		\nu^{-\alpha}$ and indicating as flat spectra those with $\alpha<$0.5.

	\section{Data reduction}
	\label{sec:reduc}
	
	\subsection{Archival data}
\begin{figure*}
		\hspace{-0.8cm}
		\includegraphics[scale=0.35]{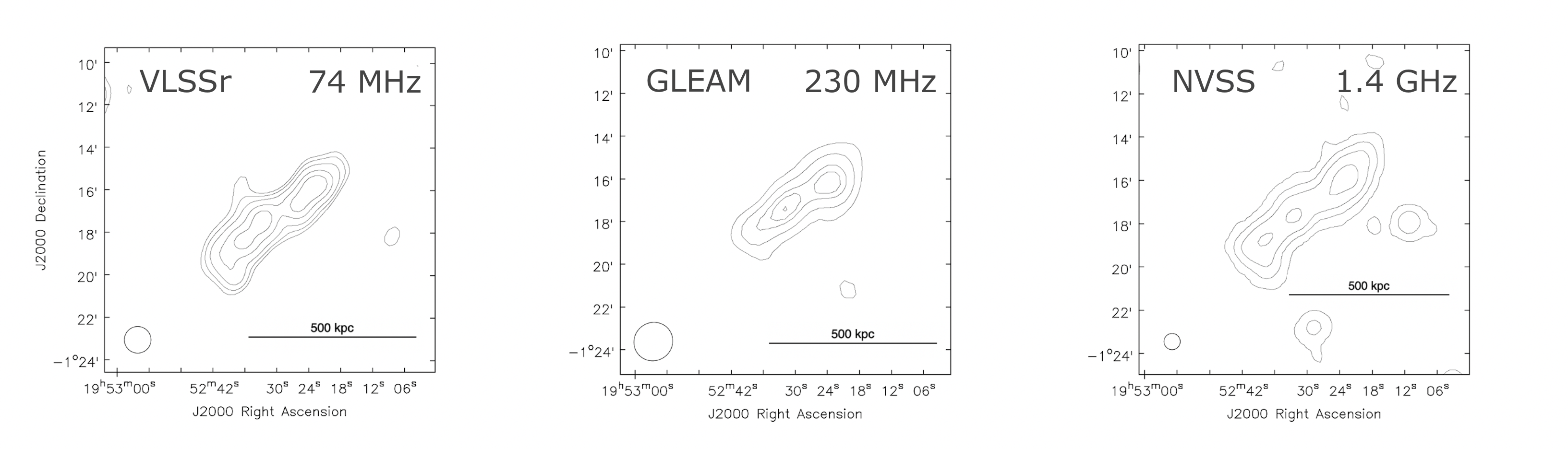}
		\caption{Radio maps of 3CR\,403.1 as seen at: (\textit{left panel}) 74 MHz (VLSSr), with contours drawn at (5, 7, 10, 15, 20)$\sigma$, (\textit{central panel}) 230 MHz (GLEAM) with contours drawn at (3, 5, 7, 8, 9)$\sigma$, and (\textit{right panel}) 1.4 GHz (NVSS) with contours drawn at (3, 30, 100, 300)$\sigma$. At all frequencies, a radio structure extending 500 kpc is detected, composed of two prominent radio lobes and core. }
		\label{radio_combo}
	\end{figure*}

	As already stated, we found a suite of available archival radio observations of 3CR\,403.1 (see Fig.~\ref{radio_combo}). In particular, this radio source was observed with the VLA as part of the NVSS in 1993, as part of the VLSSr in 2003 and with the MWA as part of GLEAM Survey in 2013. The NVSS Catalog covers the sky north of -40 deg declination ($\sim$~35000 sq. deg.). The images all have 45$\arcsec$ FWHM angular resolution and nearly uniform sensitivity, with rms brightness fluctuations approximately of 0.45 mJy beam$^{-1}$ = 0.14 K (Stokes I). The VLSSr has a resolution of 75$\arcsec$, and an average map rms noise level of $\sigma \sim$ 0.1 Jy beam$^{-1}$. 

	VLSSr covers $\sim$~30000 sq. deg. above an irregular southern boundary.
	GLEAM is an all-sky survey at 74-231 MHz, with angular resolution of 100$\arcsec$ and sensitivity between 6 and 10 mJybeam$^{-1}$,  covering a sky area of $\sim$~30000 sq. deg.

	\subsection{SRT Observations}
	\label{SRT_reduc}
	
	\begin{deluxetable*}{ccccccc}
		\tablenum{1}
		\tablecaption{Details of the SRT observations of 3CR\,403.1}\label{tab:SRT_observations}
		\tablewidth{0pt}
		\tablehead{
			\colhead{Frequency} & \colhead{Resolution} & \colhead{TOS} & \colhead{Observing date} & \colhead{OTF Mapping} & \colhead{calibrators} & \colhead{SRT Project} \\
			\colhead{(GHz)} & \colhead{(arcsec)} & \colhead{(hours)} & \colhead{} & \colhead{} & \colhead{} & \colhead{}
		}
		\decimalcolnumbers
		\startdata
		18$-$19.2        & 57         &      7           &  19-Nov-2020     & 10 R.A.$\times$10 Dec.   &  3C\,286, 3C\,84, 3C\,147           &   30-20  \\
		&            &      6           &  20-Nov-2020     & 9 R.A.$\times$9 Dec.     &  3C\,286, 3C\,84, 3C\,147     &   30-20  \\
		&            &      7           &  19-Dec-2020     & 10 R.A.$\times$9 Dec.    &   3C\,286, 3C\,84, 3C\,147    &   30-20  \\
		\enddata
		\tablecomments{Col. 1: SRT frequency range; Col. 2: SRT resolution; Col. 3: Time on source; Col. 4: Date of observation;  Col. 5: Number of images on the source; Col. 6: Calibrators; Col. 7: SRT project name.}
	\end{deluxetable*}
	
	We collected new SRT data to deeply investigate the radio components of 3CR\,403.1. The SRT is a fully-steerable 64\,m single-dish equipped with a computer-controlled active surface composed by about 1000 individual panels, that make it capable to operate with high efficiency in the frequency range from 0.3 to 100\,GHz. 
	
	We pointed at 3CR\,403.1 with the SRT K-band seven feed receiver between November, 19th and December, 19th 2020 (project ID 30-29; see Table~\ref{tab:SRT_observations} for details). We imaged a field-of-view of about $15\arcmin \times 15\arcmin$ centered on R.A.$_{\text{J2000}}$=19h52m30s and Dec.$_{\text{J2000}}$=-01d17m35s.  We acquired multiple on-the-fly scans in the equatorial frame, moving at a speed of $1\arcmin$/sec along the orthogonal R.A. and Dec. directions. The scanning speed was set as a compromise between mapping efficiency and the need to reduce the 1/f noise produced by the atmospheric and receiver gain fluctuations. 
	
	We observed in full-Stokes spectral$-$polarimetric mode in the frequency range from 18.0 to 19.2~GHz with a central frequency of 18.6\,GHz, using the SARDARA back-end \citep{melis2018}. We acquired data at a rate of about 33 spectra/sec at a spectral resolution of 1.46\,MHz. The full width at high maximum (FWHM) of the SRT beam at this frequency is 57\arcsec, and we used a pixel size of 15\arcsec~in the imaging to match the 
	transverse separation between the sub-scans. In this way, the separation of the sub-scans is equivalent to one pixel and the beam FWHM is sampled with about four independent pixels. For each pixel, we collected eight different spectra, that we averaged to increase the signal-to-noise ratio.
	
	Data reduction and imaging were performed using the SCUBE software package \citep{2016MNRAS.461.3516M}. We corrected for both the variation of telescope gain and atmosphere opacity with elevation. We flagged about 6\% of the data which were affected by the Radio Frequency Interference (RFI) both in the frequency and time-domain. The flux density was brought to the scale of \citet{2017ApJS..230....7P} using the calibrators 3C\,147 and 3C\,286. The latter was also used as absolute reference for the linear polarization position angle. We also calibrated for the on-axis instrumental polarization using the radio source 3C\,84 (J0319+4130) which is assumed to be virtually unpolarized\footnote{\url{https://science.nrao.edu/facilities/vla/docs/manuals/obsguide/modes/pol}}.
	
	We removed the baseline emission from each individual sub-scan by fitting a 2nd-order polynomial to the “cold-sky” region around 3CR\,403.1. To this aim, we masked the emission from the entire radio source using a circular region of 9.5\arcmin~in diameter, and we also use the NVSS image to identify and mask the point sources in the field of view. We then subtracted from the original sub-scan data the best-fit polynomial model we obtained from the unmasked portions of the image. In this way, we removed the unwanted contributions from the receiver noise, the atmospheric emission, and the large-scale foreground sky emission, and we retained the target emission only.
	
	We produced the spectral cubes of the full-Stokes parameters R, L, U and Q using a pixel size of 15\arcsec. To reduce the scanning noise, we combined the RA and DEC images of all the seven feeds by using the wavelet stacking algorithm described in \citet{2016MNRAS.461.3516M}. To further increase the sensitivity, we averaged all the spectral channels to produce the final images of total intensity and polarization (see Figure \ref{srt_image}).

	\subsection{\chn\  Observation}
	\label{chn_reduc}
	
	3CR\,403.1 was observed for $\sim$~8 ks with \chn\ X-ray Observatory in 2010 (see Figure \ref{combo}), as part of the 3CR \chn\ Snapshot Survey \citep{2010ApJ...714..589M} in Observation Cycle 22 (Proposal number 12700211). In the first analysis, presented in \citet{2012ApJS..203...31M}, only a marginal X-ray detection of a relatively weak radio core was reported. In \citet{2021ApJS..252...31J} the same \chn\ observation was re-analysed and X-ray extended emission on tens of kpc scale, aligned with radio emission detected only at $\sim$ 250\,MHz, was claimed at 5$\sigma$ level of significance.  
	
	We reanalyzed the \chn\ dataset and used it to carry out an X-ray spectral analysis.
\chn\ data reduction was performed using the \chn\ Interactive Analysis of Observations \citep[CIAO v4.12][]{2006SPIE.6270E..1VF} using standard procedures and threads\footnote{http://cxc.harvard.edu/ciao/threads/}, and the \chn\ Calibration Database v4.8.2. Images are all produced according to the same procedure followed for all observations of the 3CR \chn\ Snapshot Survey \cite[see e.g.,][for more details]{2015ApJS..220....5M}. Spectral analysis, in particular for the extended X-ray emission, was also performed according to methods described in our previous investigations \citep[see e.g.,][for a recent analysis]{2021ApJS..255...18M}. Here we report a brief overview of the analysis procedures.
	
	Level 2 event files were created using the CIAO task \texttt{chandra\_repro}. For the spectral analysis we used unbinned and unsmoothed X-ray images restricted to the 0.5 - 7 keV energy range to select both source and background regions. No astrometric registration was performed since the position of the X-ray nucleus is not clearly detected \citep[as already noticed in][]{2021ApJS..252...31J}. Light curves were extracted in source-free regions to check for the presence of high background intervals, that were not detected. For all analyses, blank-sky background files were used to estimate the background level at the source position. For the spectral analysis, the exposure time of the blank-sky files was adjusted so their count rates matched those of the source data in the 9.5-12 keV band \citep{2006ApJ...645...95H}.

	\section{A multiwavelength study of 3CR\,403.1}
	\label{sec:results}
	
	\subsection{MUSE overview}
	
	3CR\,403.1 was recently observed as part of the MUSE RAdio Loud Emission line Snapshot survey \citep[MURALES,][]{2019A&A...632A.124B,2021A&A...645A..12B}, thus optical spectroscopic observations of surrounding galaxies are reported here aiming at confirming the presence of a galaxy group around 3CR\,403.1. 
	
	As reported in \citet{2019A&A...632A.124B}, the line morphology in this source is particularly complex, and the observations reveal the presence of a central region ($\sim$ 9 kpc in size), with well ordered rotation and elongated structures with knots (eastern and south-eastern directions), extending up to $\sim$ 35 kpc from the galaxy nucleus, with similar velocities and redshifted by $\sim$ 150-200 km s$^{-1}$. These structures are due to ionized gas, visible as line emitting gas (mainly H$\alpha$). The spatial distribution of the ionized gas has a ``ring-like'' shape, as the other companion galaxies in the group. There are several galaxies in the MUSE field of view, but only a few emission line knots are associated with them.

	Thanks to the MUSE observations we were able to identify and estimate the redshift for several companion galaxies around 3CR\,403.1. To obtain the redshift, we fitted each line with a Gaussian, using QFitsView\footnote{https://www.mpe.mpg.de/~ott/dpuser/qfitsview.html}, which also takes into account the continuum. Spectra were extracted in each case from a circular region of 5 pixels radius, i.e., 1\arcsec\ to match the seeing of the MUSE observation, using QFitsView. We discovered five nearby companions  all marked in Figure~\ref{MUSE} together with their redshifts. We measured the velocity dispersion of the five nearby companions, obtaining a value of the sample estimate of the radial velocity dispersion to be $\sigma_v=143~\mathrm{km/s}$. According to the analysis we carried out in \citet{2020ApJS..247...71M}, assuming that this group is virialized and that galaxies have random uncorrelated velocity vectors, we estimate the total mass of the large scale environment, including dark matter, galaxies and ICM within a radius of 158 kpc (that is, the equivalent radius of the elliptical region chosen for the X-ray spectral analysis, see Section \ref{SZ}) as M$_{\text{env}}=7.5\times10^{11}M_\sun$. Since this estimate of M$_{env}$ is achieved with a low number of sources, we did not compute the statistical uncertainty. Given the mass of the environment, the size ($\sim$ 2-3 kpc) and optical luminosities ($\sim 10^9 L_\sun$, obtained from Pan-STARRS r-filter magnitudes, see \citealt{1992MNRAS.258..334S} for comparison) of the sources, this is in agreement with 3CR\,403.1 belonging to a small group of dwarf galaxies (for typical masses of a small group of dwarf galaxies see e.g. \citealt{2012AstBu..67..135M}). The mass estimate was computed using only six sources for which the signal to noise ratio of their spectra allowed us to firmly measure their redshifts (see Figure~\ref{spec} for an example of the used spectra).

	\subsection{Radio spectral analysis of the extended structure}
	\label{radio_spec}

	We first measured the flux density of 3CR\,403.1 in four radio bands (VLSSr at 74 MHz, GLEAM at 230 MHz, NVSS at 1.4 GHz and SRT at 18.6 GHz) for the entire source, and we compared these values with other global measurements taken from the literature. The global radio spectrum is determined by the sum of the spectra of both (a) the freshly accelerated electrons, whose spectrum we assume to be a power law, and (b)  the spectra of the older electron populations, whose spectrum cuts-off at high-frequency because of the radiative losses. For this reason, we make the hypothesis that the radio source is currently in the active phase, during which the radio lobes are continuously replenished with new particles by the AGN. 
	We then analysed the spatially resolved spectra at an intermediate resolution for three components: the northern lobe, the southern lobe, and the inner region in between correspondent to the radio core. All components are defined using a 3$\sigma$ limit. The flux density of the entire source was measured convolving with the VLSSr beam (75$\arcsec$), and using the 3$\sigma$ level contour from the NVSS image. For the radio components, we convolved all the images with the GLEAM beam ($\sim$ 111$\arcsec$) and then used the new 3$\sigma$ level limit of the images to define the components. We then investigated the spectral index image of the source between 1.4 and 18.6\,GHz at a slightly finer angular resolution of about 57\arcsec.  Measurements are summarized in Table~\ref{tab:fluxes} in Appendix~\ref{flux}.

	In Figure~\ref{spec}, we compare our total integrated flux measurements with literature data taken from the NASA/IPAC Extragalactic Database\footnote{https://ned.ipac.caltech.edu} and Astrophysical CATalogs support System\footnote{http://www.sao.ru/cats/} finding a good agreement.

	We fitted the integrated spectrum with a continuous injection model \citep[C.I.;][]{1970ranp.book.....P} making use of the SYNAGE software \citep[][]{1999A&A...345..769M}.
	This model is characterized by three free parameters: the injection spectral index ($\alpha_{inj}$), the break frequency ($\nu_b$), and the flux normalization. In the context of the C.I. model, it is assumed that the spectral break is due to the energy losses of the oldest relativistic electrons in the source. For high-energy electrons, energy losses are primarily due to the synchrotron radiation itself and to inverse Compton scattering on CMB photons. The spectral break marks the transition to high-frequency power law characterized by a steeper index $\alpha=\alpha_{inj}+0.5$. During the active phase, the evolution of the integrated spectrum is determined by the shift with time of $\nu_b$ to lower and lower frequencies.  Indeed, the spectral break can be considered to be a clock indicating the time elapsed since the injection of the first electron population.
	The best fit of the C.I. model yields $\nu_b$=1.9 $\pm$ 0.7 GHz.

	We estimated the equipartition field, B$_{eq}$, following the same procedure discussed in \citet{2012A&A...548A..75M}, and we obtained a value of B$_{eq}=2.4\ \mu$G, that led to a spectral age of the source of $\sim$~93 Myr, evaluated according to the equation:
	\begin{equation}
		t_{\rm syn}= 1590 \frac{B^{0.5}}{(B^2+B_{\rm CMB}^2) [(1+z)\nu_{\rm b}]^{0.5}}~\rm Myr,
		\label{synage}
	\end{equation}
	where $B$ and $B_{\rm CMB}=3.25(1+z)^2 \mu$G are the source magnetic field and magnetic field with the same energy density of the CMB, respectively, and assuming an isotropic distribution of electron pitch angles \citep[see e.g.][]{2011A&A...526A.148M,2011ApJ...729L..12M}.

	The spectra of the three components are plotted in Figure~\ref{spec} along with the fit of the JP model \citep{1974A&A....31..223J}. The JP model describes the radiative ageing of a single population of electrons with an initial power law distribution, assuming that energy losses due to the inverse Compton scattering of seed photons arising from the CMB are as relevant as radiative losses due to synchrotron emission. In these conditions, the initial power law with index $\alpha_{inj}$ develops a high-frequency exponential cut-off beyond a break frequency $\nu_{b}$. The radiative age and the break frequency are related again by Eq.\,\ref{synage}. 
	
	The spatially resolved spectral analysis, shown in Figure \ref{spec}, indicates that the spectrum of the inner parts of the radio lobes (those close to the radio source's core) is steeper than that of the outer lobes. Using basic physics, we conclude that this is due to the presence of a spectral break that shifts to lower frequencies. Indeed, in 3CR\,403.1 we find that the oldest electrons are close to the core, while we find that the electrons are younger at the tip of the lobes. This is the typical scenario found in type-2 sources \citep[see][]{1999A&A...344....7P}, where radio jets have deposited old electrons back as they make their way through the ambient gas.

\begin{figure}
		\centering
		\includegraphics[scale=0.4]{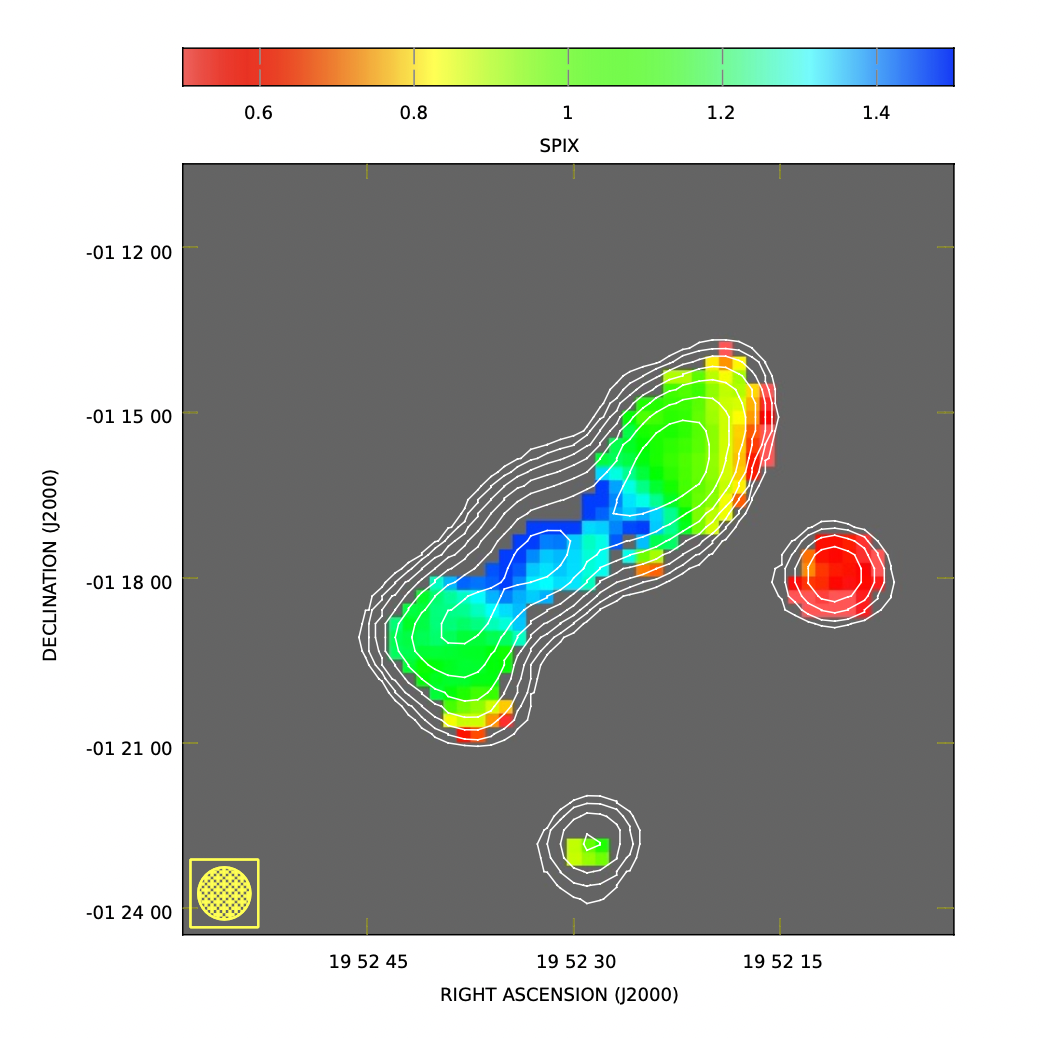}
		\includegraphics[scale=0.4]{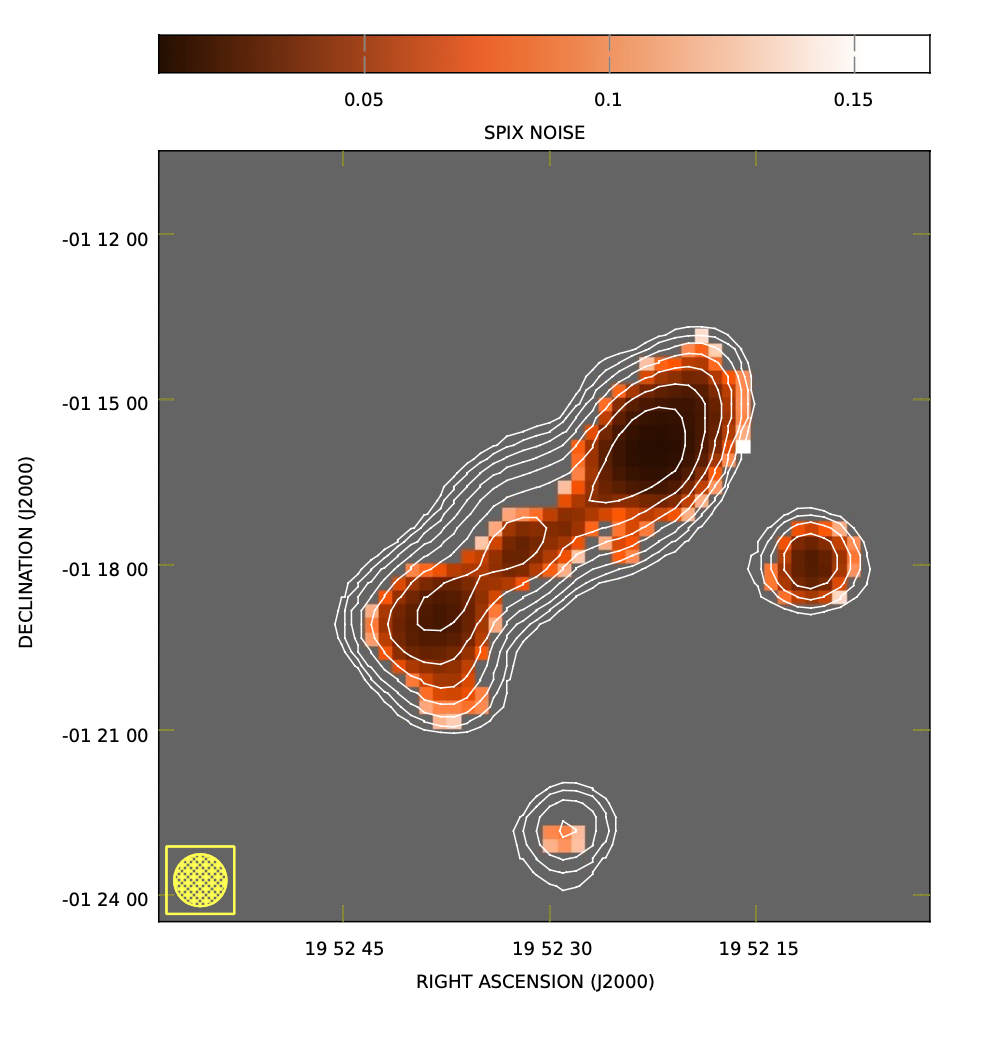}
		\caption{ \textit{Upper Panel}: Spectral index map of 3CR\,403.1 between 1.4 GHz (NVSS) and 18.6 GHz (SRT) at a resolution of  57\arcsec, as shown in the bottom left corner. The noise levels in the NVSS and
			SRT images are of 0.73 mJy/beam and 0.37 mJy/beam, respectively. The
			spectral index is calculated only in those pixels where the signal
			is above $3\sigma$ at both frequencies. White contours show the NVSS emission at 1.4 GHz and are drawn at 3$\sigma$ increasing by a factor of 2. The average statistical uncertainty on the spectral index \textit{(lower panel)} is 0.05.}
		\label{spix}
	\end{figure}

We further investigate the spectral properties of 3CR\,403.1 at a slightly higher angular resolution by analysing the spectral index map of the source between 1.4 (NVSS) and 18.6 GHz (SRT) (see Figure~\ref{spix}) according to the following steps: (1) we smoothed the NVSS image to the same resolution as that of the SRT (57\arcsec), (2) we aligned the two images using as a reference the point-like source south-west of the northern lobe and (3) we regridded the images with the new coordinates. In the spectral index map, only pixels with surface brightness above 3$\sigma$, both in the NVSS and in the SRT images, were used. Uncertainties are computed using standard uncertainty propagation formulas. We observe a steepening of the spectral index from the lobe outer edge inward to the core region, indeed confirming the characteristic trend typical of spectral type-2 sources.

	\subsection{Testing the `SZ' Signatures}
	\label{SZ}
	
\begin{figure*}
		\centering
		\includegraphics[scale=0.4]{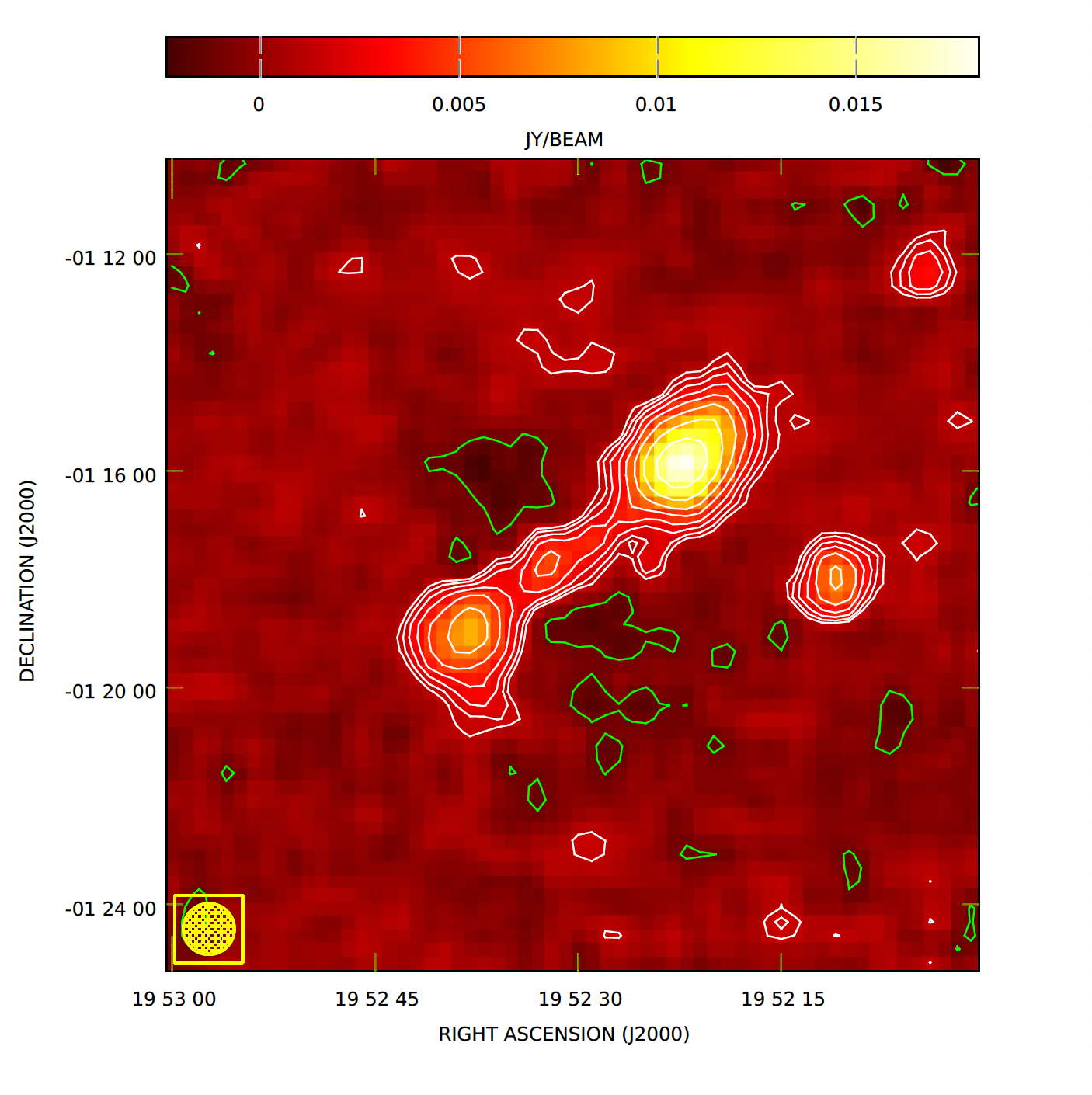}
		\caption{3CR\,403.1 total intensity image at 18.6 GHz. Contours start at 1.2 mJy/beam (3$\sigma$) and increase by $\sqrt2$. The negative green contour is traced at -1.2 mJy/beam.}
		\label{srt_image}
	\end{figure*}

	In the SRT total intensity image we noticed two regions, lying on opposite sides of the large scale radio structure, along the west-east direction, where we measure a negative intensity at 18.6\,GHz with respect 
	to the image background, as already highlighted in Figure~\ref{srt_image}.

	 We estimate that the radio intensity decrement at 18.6\,GHz is significant at more than a 3$\sigma$ level with respect to the fluctuations of the background in the SRT image (rms$=4.66\times10^{-4}$ Jy beam$^{-1}$ ).

	The negative signal in the SRT image is spatially associated with the extended X-ray
    emission revealed in the \chn\ image (see the magenta ellipse in Figure \ref{combo}). The X-ray emission is interpreted as thermal radiation from the gas of the ICM because it is not spatially associated with either GHz or MHz radio counterparts. 
   In order to carry out a deeper investigation into the origin of the negative radio signal, we compared its properties with the values we measured for the X-ray spectral parameters for the surrounding medium of 3CR\,403.1.

   	Assuming that a fraction of the radio background of the SRT observations is due to CMB, when looking in the direction of the galaxy cluster we expect a decrease of the radio intensity due to inverse Compton emission by relatively hot electrons of the ICM that upscatter the CMB radiation \citep{1972CoASP...4..173S}. Since all flux measurements are reported with respect to the background level, constituted by the CMB radiation and corresponding to the ``zero level'', we can consider the SZ effect a ``true'' negative signal in the radio images \citep{1999PhR...310...97B}. Thus we estimated the Comptonization parameter $y$ from both radio and X-ray observations, respectively, and compared them.

	At radio frequencies we computed the $y_R$ parameter starting from the change in temperature of the CMB due to the SZ effect: 
	\begin{equation}
		\frac{\Delta T}{T_{\text{CMB}}} = f(x)y_R
		\label{deltat}
	\end{equation}
	with f(x) being the frequency spectrum of the
		temperature variation and y$_{R}$ the Comptonization parameter.
	
	In the Rayleigh-Jeans approximation for the blackbody spectrum of the CMB at 18.6 GHz we have $f(x)=-2$ and then equation~\ref{deltat} becomes:
	\begin{equation}
		\frac{\Delta T_{\text{RJ}}}{T_{\text{CMB}}} = -2y_R
		\label{RJ}
	\end{equation}
	
	At radio frequencies, we use a simplified expression for the flux intensity in a given instrument beam as reported in \citealt{2016A&A...591A.142B}:
	\begin{equation}
	\left({\frac{I_\nu}{\text{mJy}/\text{beam}}}\right)
	=
	\frac{1}{340}\left({\frac{\Delta T_{\text{RJ}}}{\text{mK}}}\right)
	{\left({\frac{\nu}{\text{GHz}}}\right)}^2\left({\frac{\Omega_{\text{beam}}}{\text{arcmin}^2}}\right) 
	\label{Basu}
	\end{equation}

	Replacing equation~\ref{RJ} in \ref{Basu} we obtain:
	\begin{equation}
		y_R = - 170 \left(\frac{I_\nu}{\text{mJy/beam}}\right)
		\left(\frac{\text{GHz}}{\nu}\right)^{2}\left(\frac{\text{mK}}{T_{\text{CMB}}}\right)
		\left(\frac{\text{arcmin}^{2}}{1.13~\theta_{\text{FWHM}}^2  }\right)
	\end{equation}
	where the solid angle $\Omega_{\text{beam}}$ can be approximated as $\Omega_{\text{beam}}=1.13~\theta_{\text{FWHM}}^2$ where $\theta_{\text{FWHM}}$ is the half-power beam width.
	
	We measured the decrease of the flux intensity in an elliptical region with semiaxes  $\sim$ 125x254 kpc (see magenta ellipse in Figure~\ref{combo}) with an area equal to 0.1 Mpc$^2$, corresponding to 15.1 SRT beams, encompassing the nucleus of 3CR\,403.1 (masking the nucleus and the jets), where we detected the highest value of the negative signal. We obtained a flux intensity equal to $I_\nu$=-0.9 $\pm$ 0.1 mJy beam$^{-1}$ that corresponds to an integratex flux $S_\nu$=-21$\pm$2 mJy (in 22.3 SRT beams, assuming that the masked 7.2 beams have the same intensity). From our SRT observations we have $\theta_{\text{FWHM}}=0.95\arcmin$, and assuming T$_{\text{CMB}}$=2726 mK this leads to an estimate of $y_R=(2.0\pm0.2)\times10^{-4}$ corresponding to a value of (2.0$\pm$0.4)$\times10^{-5}$ Mpc$^2$ when measured per unit of area.

	In the X-rays the Comptonization parameter, $y_X$, is indeed related to the line-of-sight integral of the ICM pressure distribution, according to the following equation:
	\begin{equation}
		y_X=\frac{\sigma_{T}}{m_{e}c^2}\int_{L.O.S.} P_{e}(r) dl
		\label{Y_x}
	\end{equation}
	where $P_{e}(r)\sim n_{e}kT$ is the pressure profile of the thermal electrons in the ICM while $n_e$ and $T$ are the gas density and temperature, respectively.

	We analyzed the X-ray spectrum extracted from an ellipsoidal region cospatial with the flux decrement in the background CMB radiation (see Figure~\ref{combo}), adopting a thermal APEC model with Galactic absorption. We have excluded a 2\arcsec\ circular region centered on the location of the radio core where we expect to find most of the nuclear emission. Adopting for the thermal component a heavy element abundance equivalent to 0.25 of Solar, a redshift $z=0.055$ and Galactic absorption equal to 0.117$\times10^{22}$ cm$^{-2}$ as reported in \citet{2016A&A...594A.116H} we obtained a value of the temperature $kT=0.85^{+0.07}_{-0.08}$ keV and a gas density n$_{e}=(4.6\pm0.3)\times10^{-4}$cm$^{-3}$. The mass of the gas in this ellipsoidal volume estimated as reported in \citet{messias2021}  is equal to M$_{gas}\simeq(2.2\pm0.1)\times10^{11} M_\sun$.

	We estimated $y_X$ using the values of temperature and density obtained from the spectral fit in the ellipse.
	With $P_{e}\sim n_{e}kT$ we obtained $P_{e}=6.11^{+0.63}_{-0.69}\times10^{-13}$ erg cm$^{-3}$. Assuming a constant value for $P_{e}$, from Equation~\ref{Y_x}  we obtained $y_X=3.82^{+0.40}_{-0.43}\times10^{-7}$, corresponding to a value of $3.82^{+0.40}_{-0.43}\times10^{-8}$ Mpc$^2$ when measured per unit area.

	The two values we obtained from the X-ray and radio analysis, are not in agreement, with a discrepancy of three orders of magnitude.
	
	The thermal energy of a gas can be computed starting from the ideal gas law as:
		E$_{\text{th}}$=$\frac{3}{2}$nKTV where in this case n and T are respectively the density and temperature obtained from the X-ray spectral analysis in the ellipsoidal volume V=4.9$\times10^{71}$ cm$^{3}$.
		This expression results in E$_{\text{th}}$=4.46$^{+0.46}_{-0.50}\times10^{59}$ erg.
		We can calculate the thermal energy also making use of the Comptonization parameter as described in equation~\ref{Y_x} using for y the value obtained in the radio (y$_{R}$=(2.0$\pm$0.2)$\times10^{-4}$).
		This results in
		E$_{\text{th}}$=$\frac{3}{4}y_R\frac{m_{e}c^2}{\sigma_{T}}\frac{V}{R}$ where V is again the volume of the ellipse and R=3.9$\times10^{23}$ cm is the semiaxe of the ellipse used to compute the y$_R$ parameter. In this case we obtain E$_{\text{th}}$=(2.34$\pm0.23)\times10^{62}$ erg. We observe the same discrepancy obtained for the radio-X-ray Comptonization parameters.

	We indeed considered the possibility that the negative bowl in the radio image is an artifact left from the data reduction process. We ran numerous  tests to determine if the radio diminution could be considered an artifact or not, such as comparison of the first and last SRT observations, baseline subtraction, more tailored masking, but the results are not conclusive (see details in Appendix \ref{test}). 
	
	If we assume that the negative signal is not affected by artifacts, we can then make the hypothesis that the discrepancy of the radio and X-ray values of the Comptonization parameter could be partially due to the presence in the ICM around 3CR\,403.1 of a pool of non thermal, mildly relativistic ``ghost'' electrons, produced during a past activity of the radio source. We assume that the electrons should be mildly relativistic otherwise a thermal bath of electron would have a same total thermal energy of a relatively massive group or cluster, making this unreliable. These electrons have energy to radiate at such low radio frequencies that cannot be detected but they could still scatter the CMB photons causing the observed negative signal around 3CR\,403.1 at 18.6\,GHz. Another hint to the AGN past activity is also the presence of the radio knots seen in the seen in an archival VLA P-band observation (not shown), in the same direction of the negative radio signal.
	
	To summarize, there are many possibilities for this decrement (real and artifacts): 1) SZ effect from ICM; 2) imaging artifacts; 3) a pool of non thermal, mildly relativistic ``ghost'' electrons emitted during a past activity of the radio source, that then cooled and can partially explain the discrepancy of the X-ray and radio Comptonization parameter values (see Appendix \ref{test} for more details). 
	We also highlight that this is the first time a similar effect has been detected in a radio galaxy at such low redshift, thus deeper observations are needed to investigate this effect in more detail.

	\subsection{Magnetic Field measurements}
	\label{Bfield}
	
	We used SRT data combined with NVSS to derive the rotation measure (RM) image of 3CR\,403.1 (see Figure~\ref{rm}). The RM provides information on the intracluster magnetic fields, since a magnetized plasma changes the properties of the polarized emission coming from a radio source embedded in (or in the background of) the galaxy group/cluster. The position angle of the linearly polarized radiation rotates by an amount that is proportional to the integral of the magnetic field along the line-of-sight, times the electron density of the ICM (Faraday rotation effect, \citealt{1979AJ.....84..725D}). 
	
	In the case of an external Faraday screen, i.e. if the magnetoionic medium is located 
	between the radio source and the observer, the polarization angle rotates according to the $\lambda^2$-law:
	\begin{equation}
		\psi_{obs}(\lambda)=\psi_{Int}+\lambda^2 RM
		\label{RM}
	\end{equation}
	where $\psi_{obs}$ is the observed polarization at wavelength $\lambda$ while $\psi_{Int}$ is the intrinsic polarization angle
	\citep[see e.g.][]{2004IJMPD..13.1549G}.
	
\begin{figure*}
		\hspace{-0.5cm}
		\includegraphics[scale=0.5]{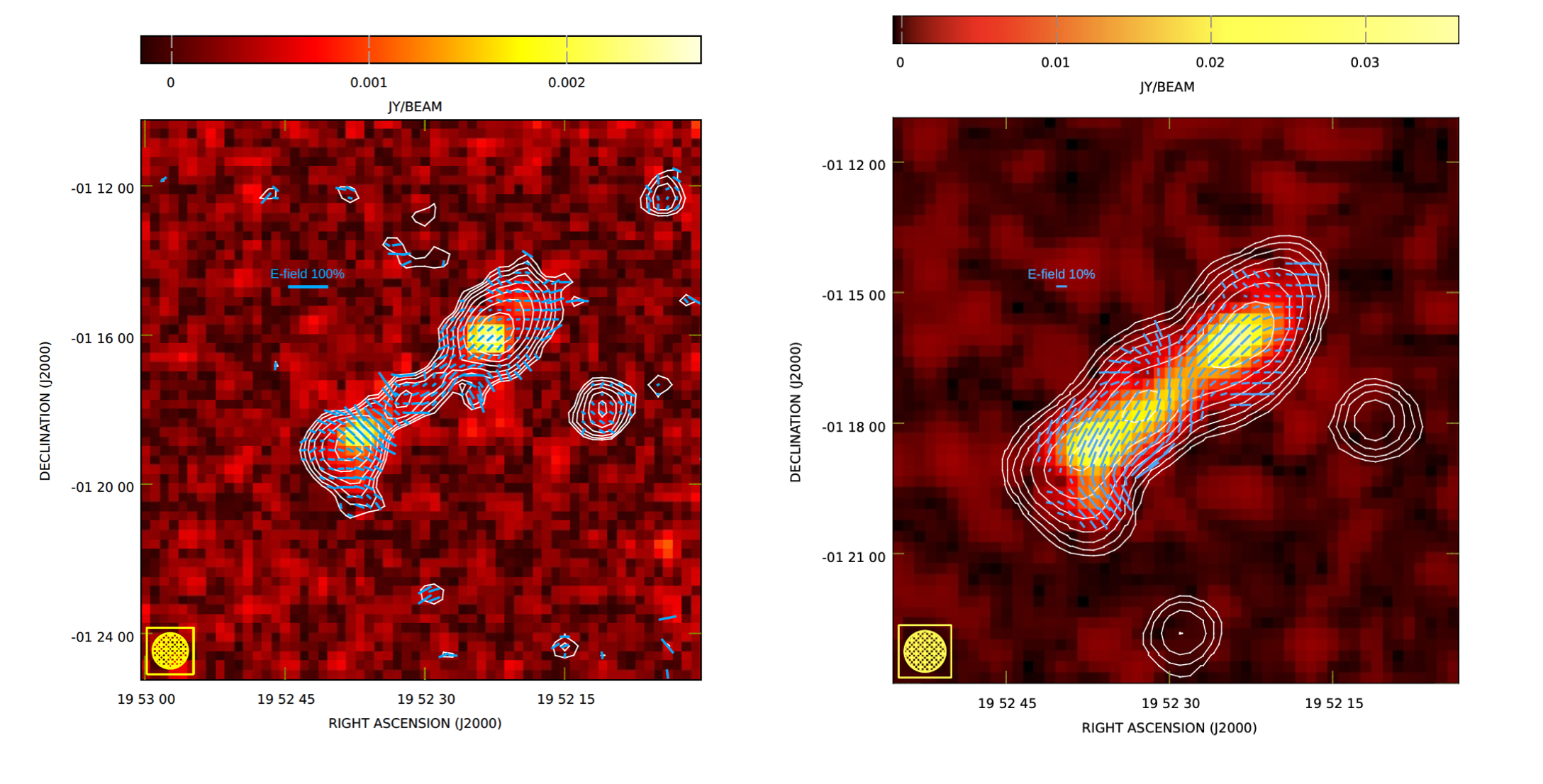}
		\caption{\textit{Left panel}: 3CR\,403.1 polarization image at 18.6\,GHz (SRT). The contours refer to the total intensity.  The polarization vectors represent the direction of the radio wave E-field, with their length proportional to the fractional polarization, while their orientation represents the polarization angle. The average fractional polarization of 3CR\,403.1 at 18.6 GHz and 57\arcsec~resolution is FPOL=0.17. \textit{Right panel}: Polarized intensity image of 3CR\,403.1 at 1.4 GHz (NVSS). The contours refer to the total intensity. The polarization vectors represent the direction of the radio wave E-field. Their length is proportional to the fractional polarization while their orientation represents the polarization angle.}
		\label{pol}
	\end{figure*}

	By measuring the angle at different frequencies it is possible to derive the RM by a linear fit to Eq.~\ref{RM}. In our case we simply measured the difference $\Delta \psi$ in the polarization angle between the SRT and the NVSS images (see Figure~\ref{pol}) and we computed the RM as:
	\begin{equation}
		RM=\Delta \psi/(\lambda_1^2-\lambda_2^2) \qquad \rm rad/m^2
		\label{RMformula}
	\end{equation}
	where $\lambda_1$= 0.016\,m and $\lambda_2$=0.21\,m are the wavelengths corresponding the SRT and NVSS images.

	The RM image is shown in Figure~\ref{rm} and has been derived considering only those pixels where the uncertainty in the polarization angle
	is less that 10$^{\circ}$ and the total intensity is above the 3$\sigma$ level at both frequencies. The SRT data provide a polarization angle very close to $\psi_{Int}$, since the rotation is negligible at such a high frequency for typical cluster RMs. However, since we have only two measurements it is not possible to resolve possible $n\pi$-ambiguities on the observed polarization angle in the NVSS image. We observe a net asymmetry in the RM of the two lobes.  As shown in Figure~\ref{rm}, the southern lobe has an average $RM \simeq -26$ rad/m$^{2}$ while for the northern lobe we observe an average Faraday rotation as low as $RM \simeq 1$ rad/m$^{2}$.

	We modeled the observed RM asymmetry assuming that 3CR\,403.1 is located at the center of the galaxy group and is inclined with respect to the plane of the sky so that the southern
	lobe has a larger Faraday depth in comparison to the northern lobe \citep[see e.g.][]{2008MNRAS.391..521L}.
	
	We considered an idealized situation in which the magnetic field is composed by 
	uniform ``cells'' of size $\Lambda_{B}$ with random direction in space and we used the software FARADAY \citep{2004A&A...424..429M} to calculate the variance of the Faraday rotation ($\sigma^{2}_{\rm RM}$) from a depth $L$ and a projected radius $r_{\perp}$:
	\begin{equation}
		\sigma^{2}_{\rm RM}(L) =812^2 \Lambda_{B} \int_{L}^{+\infty}(n B_{\parallel})^2  dl  ~~~({\rm rad^2m^{-4}})
		\label{felten}
	\end{equation}
	where the cluster's midplane is located at $L=0$, the cluster far side is located at $L<0$, and the cluster near side is at $L>0$ \citep[e.g.][]{1982ApJ...252...81L,1991MNRAS.250..726T,1995A&A...302..680F,1996ASPC...88..271F}. In Eq.\,\ref{felten}, the magnetic field strength is in $\mu G$, the density in $cm^{-3}$, while the field scale and the physical depth are in kpc. 
	
	If the electron gas density is described by the $\beta$-model \citep[][]{1976A&A....49..137C}:

\begin{equation}
	    n=n_{0}\left(1+\frac{r^{2}}{r_{c}^{2}}\right)^{-\frac{3}{2}\beta},
	\end{equation}
	
	with r$_{c}$ being the core radius of the gas distribution and we assume that the magnetic field strength scales with the gas density according to:
		\begin{equation}
	    B=B_0(n/n_0)^{\eta}
	    \label{Bscaling}
	\end{equation}
	
	then, by substituting in Eq.\,\ref{felten}, we find
	\begin{eqnarray}
		&&\sigma^{2}_{\rm RM}(L, r_{\perp}) = 812^2 \Lambda_{c} n _0^2B_0^{2}/3\times \nonumber \\
		&&\times \int_{L}^{+\infty}(1+r_{\perp}^2/r_c^2+l^2/r_c^2)^{-3\beta(1+\eta)}dl  ~~~{\rm rad^2m^{-4}}
	\end{eqnarray}
	where $r_{\perp}$ is the impact parameter from the cluster centre, and we assume that the magnetic field is isotropic so that $B_{\parallel}=B/\sqrt{3}$.
	
	For the $\beta$-model we assumed a core radius $r_c=200$ kpc and $\beta=0.6$, while for the magnetic field model we assumed a correlation length of $\Lambda_B=20$ kpc and $\eta=1$. These are the best fit parameter found for A\,194 by \citet{2017A&A...603A.122G}. The central gas density in A\,194 is similar to that of 3CR\,403.1, and we speculate that the properties of the intracluster magnetic field derived for A\,194 using high-resolution RM data can be assumed as a reference also for the case of 3CR\,403.1. Indeed, we fixed $\Lambda_B$ and $\eta$, and we deduce the combinations of the central magnetic field strength, $B_0$, and the source inclination with respect to the plane of the sky that explain the observed RM asymmetry between the two lobes.
	
	In the upper right panel of Figure~\ref{rm} we show the expected RM profiles as a function of the
	depth for three values of the magnetic field strength at the galaxy group centre. The dots represent the measured RM values for the south and north lobe if an inclination of $60^{\circ}$ is assumed. For the southern lobe we considered a projected distance of $r_{\perp}$=119 kpc while for the northern lobe we considered $r_{\perp}$=194 kpc. In the bottom right panel of Figure~\ref{rm} we show the RM difference between the two lobes versus the source's inclination. The shaded horizontal stripe represents the measured RM difference. We do not have any a priori hint about the source inclination. However, in the 68\% of the cases (inclination $> 30^{\circ}$) the observed RM can be used to constrain the cluster magnetic field strength, $B_0$, between 1.75 and 3.5 $\mu$G. This is the magnetic field at the cluster centre. Note that according to Eq.\,\ref{Bscaling} with $\eta=1$, the field strength decreases with radius following the gas density.
	
\begin{figure}
		\centering
		\hspace*{-1cm}
		\includegraphics[scale=0.23]{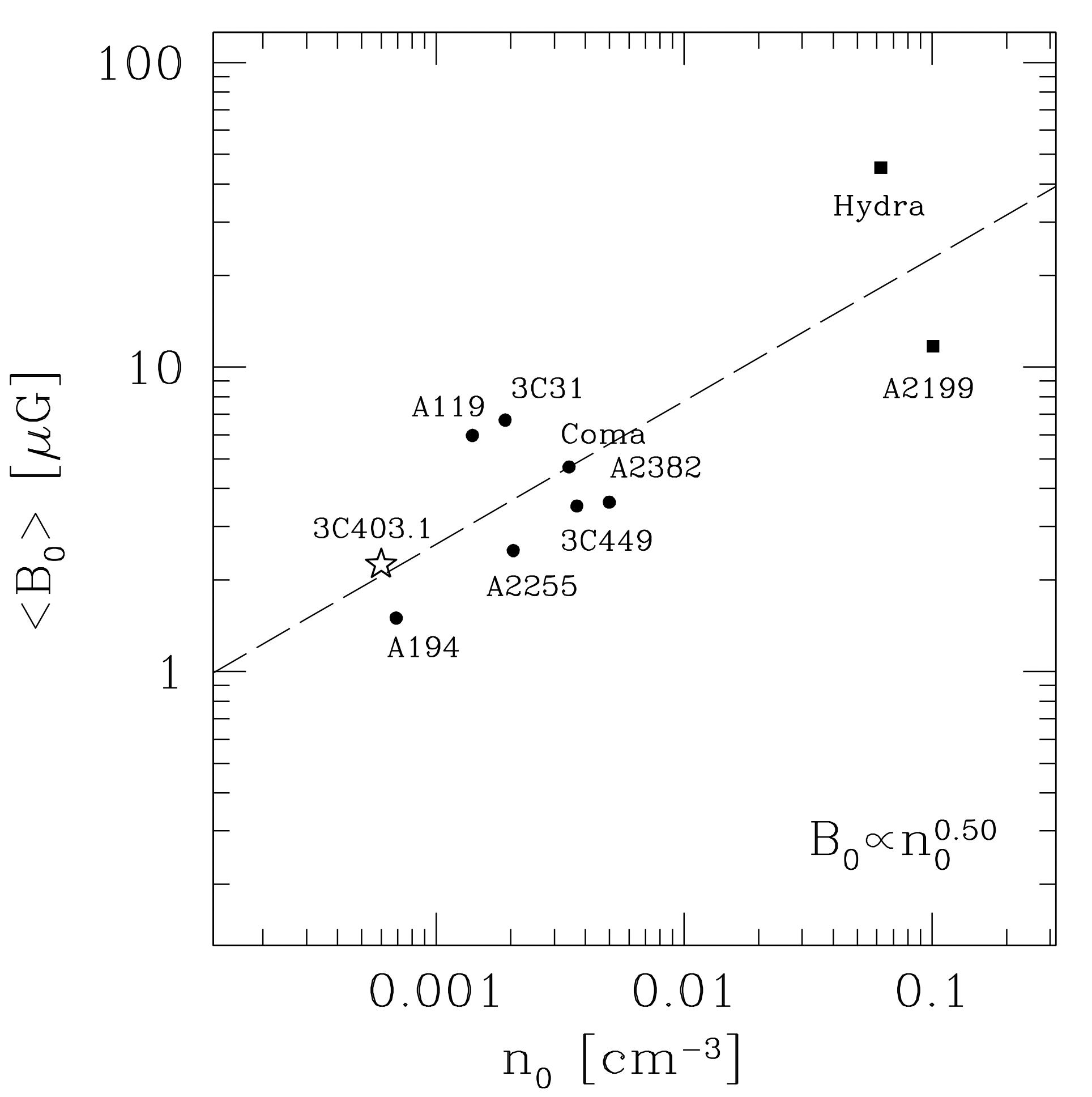}
		\caption{Central magnetic field strength $<B_0>$ vs the central electron density n$_0$. The dashed line indicates the scaling obtained by a linear fit of the log-log relationship $<B_0>\propto n_{0}^{0.50}$. 3CR\,403.1 is indicated with a star. The other sources in the plot are discussed in \citealt{2017A&A...603A.122G}.}
		\label{Bdens}
	\end{figure}

	In Figure~\ref{Bdens} we plot 3CR\,403.1 in the $B_0$-density plane from \citet{2017A&A...603A.122G}. The position of the radio source in the plane is consistent with the correlation observed for other galaxy clusters for which the magnetic field strength was determined from the RM analysis. The low central magnetic field found in this work for the poor galaxy cluster around 3CR\,403.1 confirms the general trend that fainter central magnetic fields seem to be present in less dense galaxy clusters \citep{2017A&A...603A.122G}.

	\section{Summary and conclusions}
	\label{sec:conclusion}
	
\begin{figure}[!]
		\hspace{-1cm}
		\centering
		\includegraphics[scale=0.4]{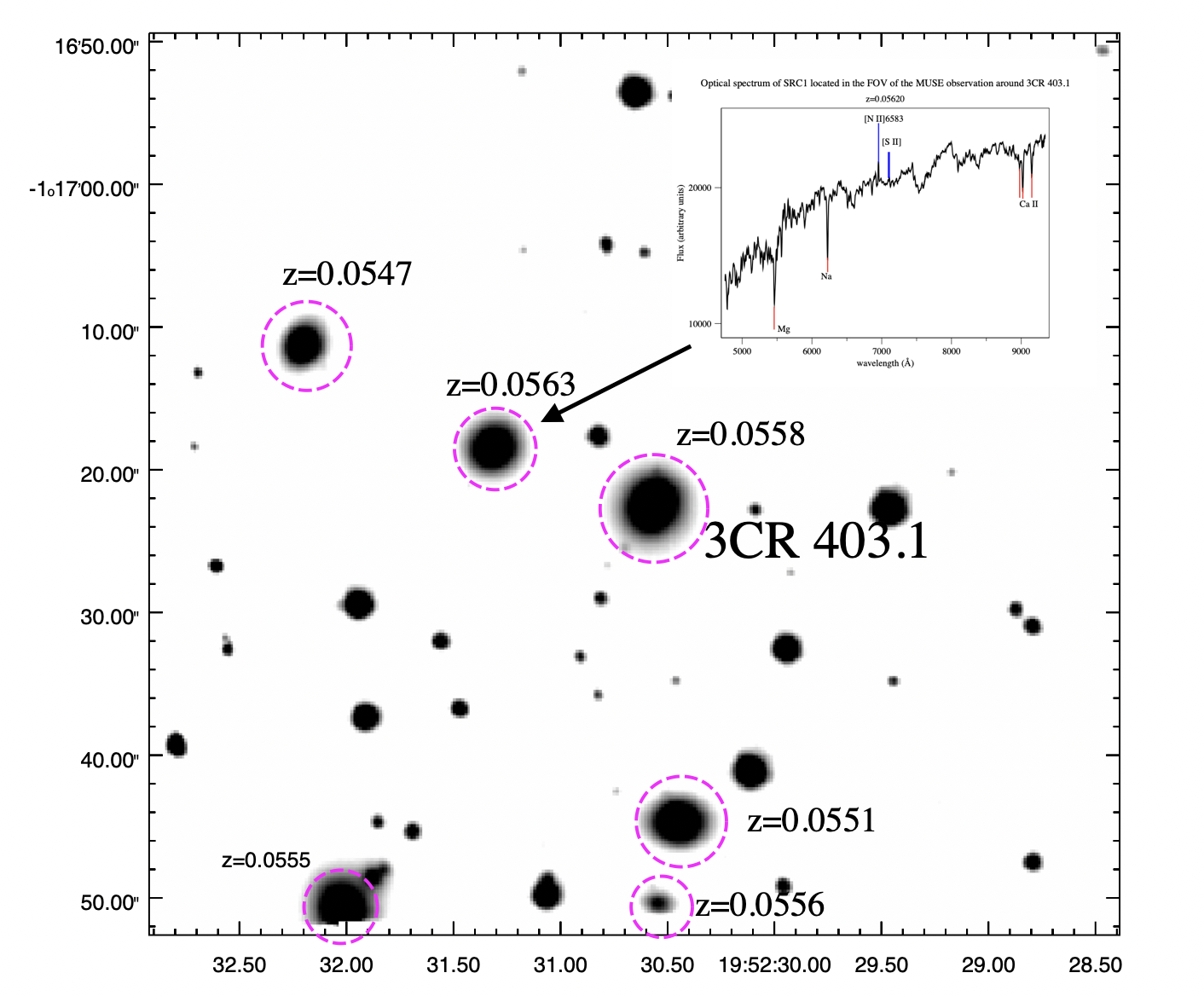}
		\caption{MUSE white-light observation of the 3CR\,403.1 field (1\arcmin~square or 4168 kpc$^2$). Sources in the same group as our target are indicated with magenta circles. For the other sources in the field, an estimate of $z$ was not possible due to a low signal-to-noise ratio. For one of the sources in the group, the spectrum used to estimate the redshift is added.}
		\label{MUSE}
	\end{figure}

	We have presented a spectral-polarimetric study of the FR\,II radio galaxy 3CR\,403.1, hosted in a small galaxy group of dwarf galaxies, as shown in the observations performed by VLT/MUSE as part of the MURALES survey (see Figure~\ref{MUSE}). This study was carried out using new high frequency radio data from the SRT and archival data from VLA and MWA. Radio data were complemented with X-ray data from \chn , obtained as part of the 3CR Snapshot Survey \citep[see also recent observations in][]{2018ApJS..234....7M,2018ApJS..235...32S,2020ApJS..250....7J} and optical observations available from the MURALES survey.
	
	Assuming that this group is virialized, we used the velocity dispersion of the five companion galaxies detected with MUSE to estimate the total mass of the system as $M_{env}=7.5\times10^{11}M_\sun$.
	
\begin{figure}
		\hspace*{-1cm}
		\centering
		\includegraphics[scale=0.45]{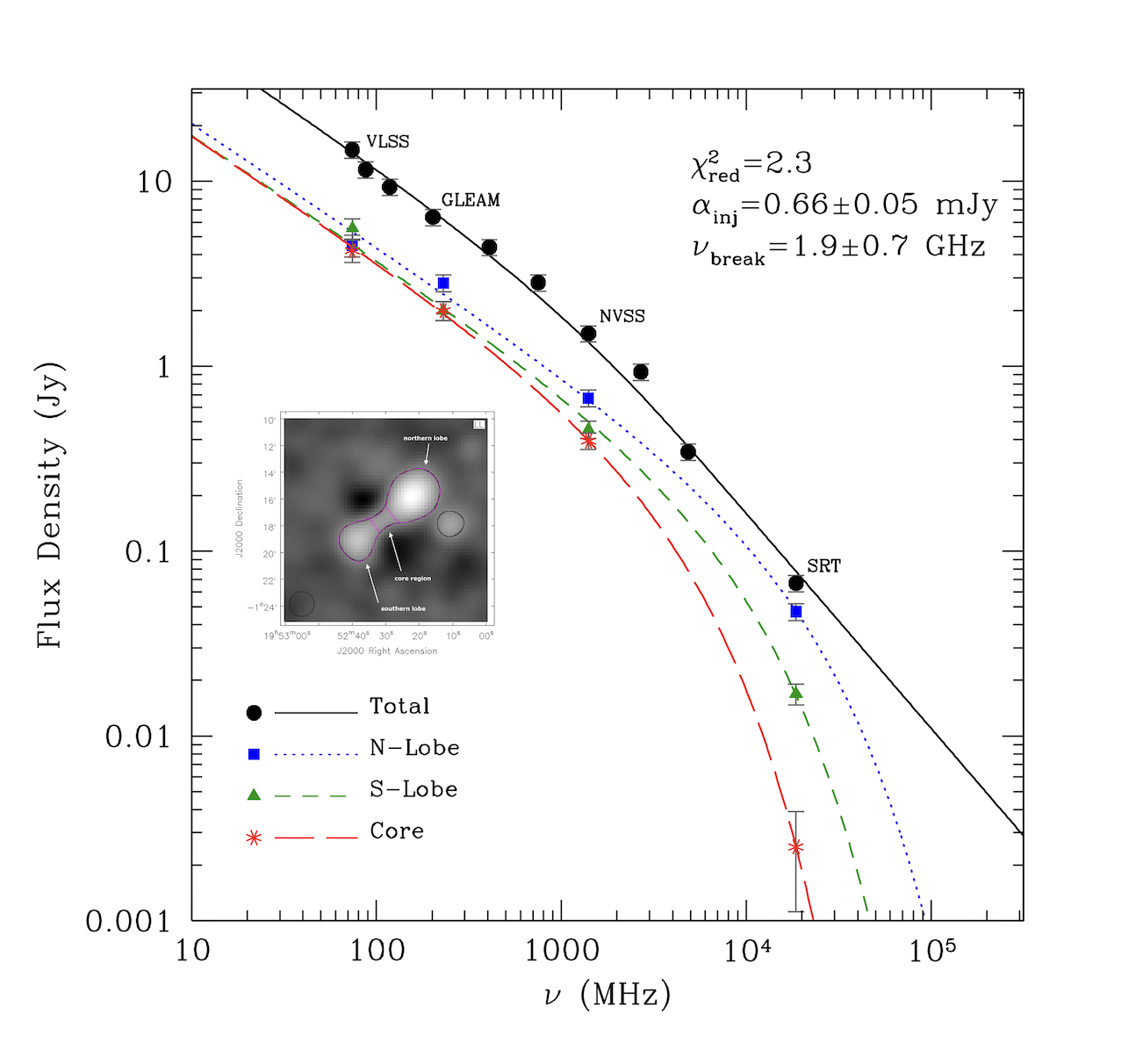}
		\caption{Total radio spectrum of 3CR\,403.1. Black circles represent the flux densities measured in this work, complemented with values taken from the literature. The continuous black line is the best-fit of the CI model and shows the presence of a spectral break at a frequency of 1.9\,GHz, followed by a moderate steepening. Blue squares, green triangles, and red stars represent the spectra of the north lobe, the south lobe, and the core regions respectively (highlighted in the image added on the plot). We modelled these components  using a Jaffe-Perola whose best-fit is represented by the blue dotted, green short-dashed, and red  long-dashed lines. See Section \ref{radio_spec} for more details.}
		\label{spec}
	\end{figure}

	We measured the flux density of 3CR\,403.1 for the entire source and its components, radio core region and lobes, separately, using new SRT observations, complemented with archival NVSS, GLEAM and VLSSr radio data (lower frequencies). Results of the spectral analysis are shown in Figure~\ref{spec}.
	
	Given the values of the parameters obtained from the fit of the radio spectrum for the entire radio source (injection spectral index and break frequency), we measured the equipartition magnetic field, finding a value of B$_{eq}=2.37\ \mu$G. This value was used to compute the spectral age of the source, being 93 Myr.
	
\begin{figure}
		\centering
		\includegraphics[scale=0.2]{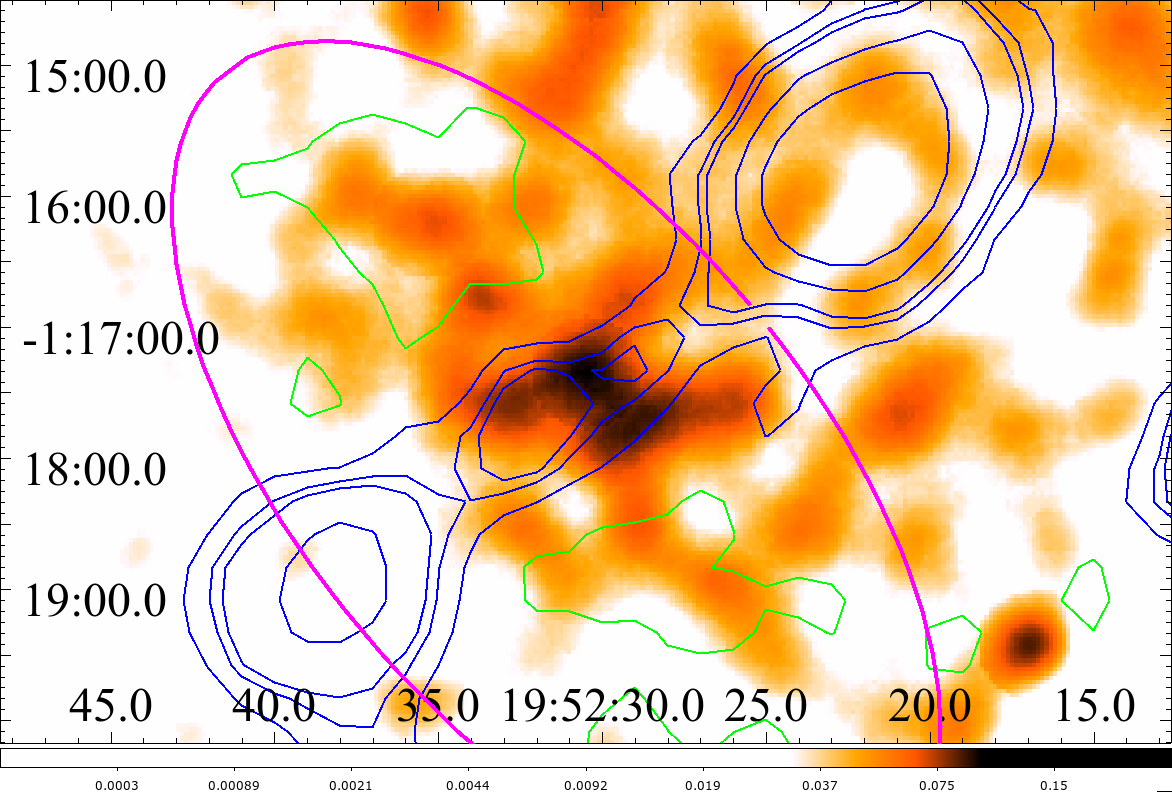}
		\caption{\chn\ X-ray image of 3CR\,403.1, filtered in the 0.5–3 keV band, binned up to 1.968\arcsec~pixel$^{-1}$ and smoothed with a 25.6\arcsec~Gaussian kernel. Radio contours (SRT, 18.6\,GHz) are drawn at (3, 5, 6, 10, 15, 30, 50)$\sigma$. The negative green contour is traced at -1.2 mJy/beam (see Figure~\ref{srt_image}). The magenta ellipse indicates the region used to extract the X-ray spectrum. This region is cospatial with the flux decrement observed in the SRT image. See Section \ref{chn_reduc} for more details.}
		\label{combo}
	\end{figure}

	From the new high-frequency SRT data at 18.6\,GHz we unexpectedly observed a flux depression in two regions perpendicular to the radio axis (see green contours in Figure~\ref{combo}) with a level of significance higher than 3$\sigma$. This negative radio signal is cospatial with the extended X-ray emission detected in the \chn\ observation, that we interpreted as thermal radiation from the hot ICM surrounding 3CR\,403.1.

	Thus, we re-analysed the \chn\ observation, focusing on the region of the radio flux depression (see Figure~\ref{combo}). We performed a spectral analysis of the X-ray emission to estimate the temperature and density of the ICM in that region. We found $kT=0.85^{+0.07}_{-0.08}$ keV and n$_{e}=(4.6\pm0.3)\times10^{-4}$cm$^{-3}$ that corresponds to a gas mass equal to M$_{gas}\simeq(2.2\pm0.1)\times10^{11} M_\sun$.

	The radio Comptonization parameter would require a much larger ICM pressure than that estimated from the X-rays and therefore we investigated the possibility that  the negative signal in the SRT image is an artifact, but the results were not conclusive.
	We obtained an estimate of the Comptonisation parameter both in the radio ($y_R=(2.0\pm0.4)\times10^{-5}$ Mpc$^2$) and in the X-ray ($y_X=3.82^{+0.40}_{-0.43}\times10^{-8}$ Mpc$^2$) from an elliptical region encompassing the nucleus of 3CR\,403.1 where we detected the highest value of the negative signal. From these calculations we find that the Comptonization parameter values obtained with the radio (y$_R$) and X-ray (y$_X$) are not consistent. We investigated the possibility that the negative signal in the SRT image is an artifact, but the results were not conclusive. If the negative signal is considered as real, than this implies that (i) the X-ray observations are not deep enough to allow us to make a more accurate estimate or (ii) we speculate that part of the discrepancy of the two values is due to the presence of a pool of non-thermal, mildly relativistic old electrons,  ejected from the radio source's core in a past episode of activity (as suggested by a VLA P-band observation). 
	
	Due to the low statistics of the available X-ray data, no firm estimate of the Comptonization parameter can be drawn. We plan to collect more radio data to investigate the emission detected by the VLA in P band, possibly related to past AGN activity and thus the presence of cold electrons responsible of the flux decrease observed in SRT data.
	
\begin{figure*}
		\includegraphics[scale=0.4]{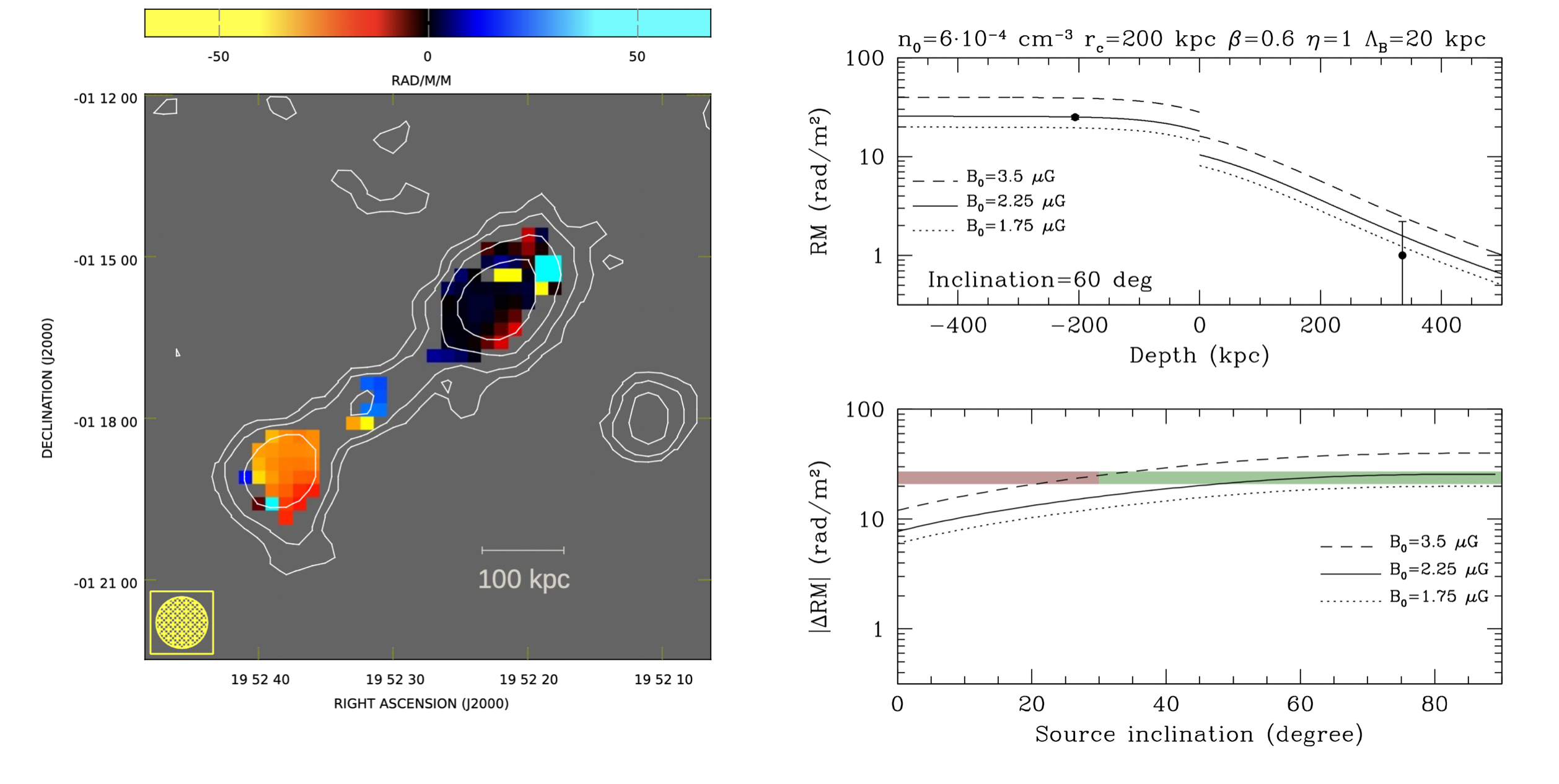}
		\caption{\textit{Left Panel}: Rotation measure image calculated with the images at 1.4 and 18.6 GHz smoothed with a FWHM Gaussian of 57\arcsec. The color range is from -60 to 60 rad/m$^2$. Contours refer to the total intensity image at 18 GHz. Levels start at $3\sigma$ and increase by a factor of 2.\textit{Right panel, top}: absolute value of the RM as a function of the physical depth of the radio lobes in the cluster medium. The dots represent the measured value for the south (left) and north (right) lobes, plotted at a depth corresponding to a source inclination of $60^{\circ}$. The three lines represent the expected RM signal for three different values of magnetic field strength at the cluster centre. \textit{Right Panel, bottom}: absolute value of the RM difference between the two lobes versus the source inclination. The shaded region represents the measured value, while the lines are the expected RM difference for three different values of magnetic field strength at the cluster centre.}
		\label{rm}
	\end{figure*}

	Finally, we used SRT data combined with NVSS to derive the rotation measure (RM) images of 3CR\,403.1 (see Figure~\ref{rm}). From this estimate we concluded that the ICM surrounding 3CR\,403.1 is permeated by a magnetic field with strength between 1.75 and 3.5 $\mu$G. Thus, we compared this value of the magnetic field with the density found from the X-ray analysis. The position of the radio source in the plane is consistent with the correlation observed for the other galaxy clusters for which the magnetic field strength was determined from the RM analysis.
	
	We want to stress that the SRT observations of 3CR\,403.1 gave intriguing results, but this should be a starting point for an exploratory program of powerful radio sources looking for similar effects, opening a new window on the scientific results achievable with the SRT.

	\begin{acknowledgments}
		
		We thank the referee for a careful reading of our manuscript and many helpful comments that led to improvements in the paper.
		This work is supported by the ``Departments of Excellence 2018 - 2022'' Grant awarded by the Italian Ministry of Education, University and Research (MIUR) (L. 232/2016).
		The Sardinia Radio Telescope is funded by the Ministry of University and Research (MIUR), Italian Space Agency (ASI), and the Autonomous Region of Sardinia (RAS) and is operated as National Facility by the National Institute for Astrophysics (INAF). This research has made use of the CATS database, operated at SAO RAS, Russia \citep{2005BSAO...58..118V}. This research has made use of the NASA/IPAC Extragalactic Database (NED) which is operated by the California Institute of Technology, under contract with the National Aeronautics and Space Administration. A.P. acknowledges financial support from the Consorzio Interuniversitario per la Fisica Spaziale (CIFS) under the agreement related to the grant MASF\_CONTR\_FIN\_18\_02.  A. J. acknowledges the financial support MASF\_CONTR\_FIN\_18\_01 from the Italian National Institute of Astrophysics under the agreement with the Instituto de Astrofisica de Canarias for the ``Becas Internacionales para Licenciados y/o Graduados Convocatoria de 2017’’.  W.F. and R.K. acknowledge support from the
		Smithsonian Institution and the Chandra High Resolution Camera Project through NASA contract NAS8-03060.  W.F. acknowledges additional support from NASA Grants 80NSSC19K0116, GO1-22132X, and GO9-20109X.  
	\end{acknowledgments}
	
	\vspace{5mm}
	\facilities{\chn,\ MWA, SRT, VLA, VLT/MUSE.}
	
	\software{CASA \citep{2007ASPC..376..127M}, CIAO \citep{2006SPIE.6270E..1VF}, FARADAY \citep{2004A&A...424..429M}, SCUBE \citep{2016MNRAS.461.3516M}, Sherpa \citep{2001SPIE.4477...76F, 2007ASPC..376..543D, 2020zndo...3944985B}, SYNAGE \citep{1999A&A...345..769M} }.

	\clearpage

	\appendix	
	\section{Radio fluxes}	
	\label{flux}
	
	In this appendix we report the flux densities of 3CR\,403.1 and of its radio components, using GLEAM convolved beam.
	
	\begin{deluxetable*}{ccccc}[h!]
		\tablenum{1}
		\tablecaption{Flux densities of 3CR\,403.1 and of its morphological features. Fluxes of core and lobes regions have been measured using GLEAM convolved beam.}
		\label{tab:fluxes}
		\tablewidth{0pt}
		\tablehead{
			\colhead{Frequency} & \colhead{Flux density core region} & \colhead{Flux density southern lobe} & \colhead{Flux density northern lobe} & \colhead{Flux density entire} \\
			\colhead{(GHz)} & \colhead{(mJy)} & \colhead{(mJy)} &  \colhead{(mJy)} & \colhead{(mJy)}
		}
		\startdata
		0.074   &    4210$\pm$580  &   5560$\pm$690   &   4500$\pm$610    &   14800$\pm$200\\
		0.230 & 2000$\pm$230  &   2000$\pm$230  &   2820$\pm$300  &    3490$\pm$100    \\
		1.4 &   394$\pm$40   &    455$\pm$50   &   672e$\pm$70   &   1500$\pm$1       \\
		18.6 &  2.51$\pm$1.39    &   16.9$\pm$2.2    &   47$\pm$5   & 67.4$\pm$0.5 
		\enddata
		
		\tablecomments{Col. 1: Frequency; Col. 2: Flux density of the core region; Col. 3: Flux density of the southern lobe; Col. 4: Flux density of the northern lobe;  Col. 5: Flux density of the entire source.}
	\end{deluxetable*}
	
	\section{SRT background tests}
	\label{test}
\begin{figure*}
	\hspace*{-1cm}
	\includegraphics[scale=0.54]{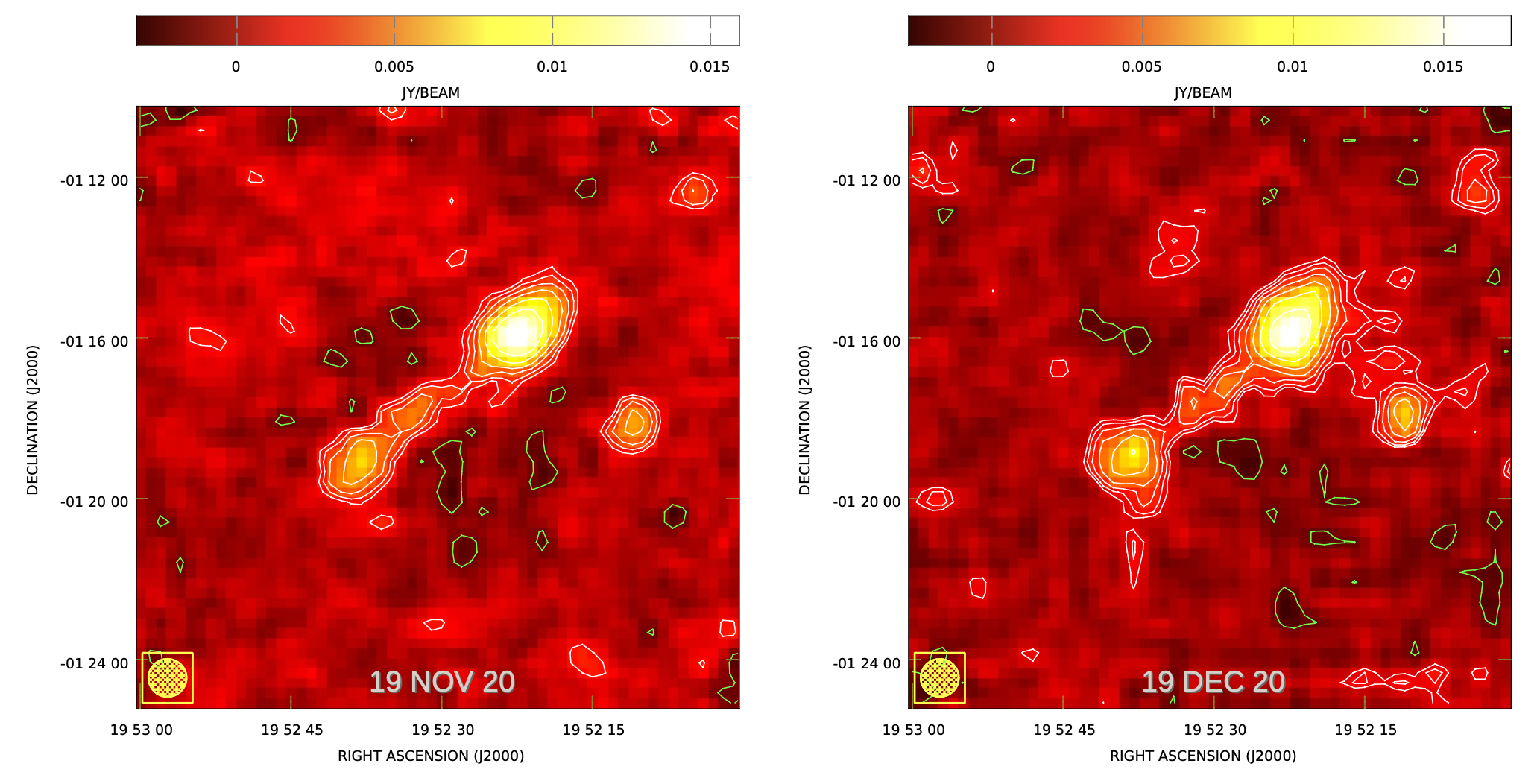}
	\caption{Comparison of Nov 19th (\textit{left panel}) versus Dec 19th, 2020
(\textit{right panel}) session. The rms noise levels are 0.60 and 0.52 mJy/beam,
respectively. Contours are traced at -3$\sigma$, 3$\sigma$ and scale by $\sqrt{2}$.
Negative contours are represented in green.}
	\label{19nov_vs_19dec}
\end{figure*}

	We analyzed separately the images of the two observing days 19 Nov and 19 Dec 2020, which have comparable noise levels (see Figure~\ref{19nov_vs_19dec}). Although the signal-to-noise ratio is lower than that of the combined image, there is a hint that the negative signal is present in both the individual images. This would exclude that it is an artifact due to peculiar atmospheric fluctuations not captured by the baseline subtraction.

	We then repeated the baseline subtraction by i) using a linear fit instead of a 2nd-order polynomial fit and ii) by dropping the mask completely and using just the 10\% of data from the begininning and end of each individual sub-scan to define the``cold-sky". In both the cases the negative signal around the source is still observed.
	
\begin{figure*}
		\includegraphics[scale=0.67]{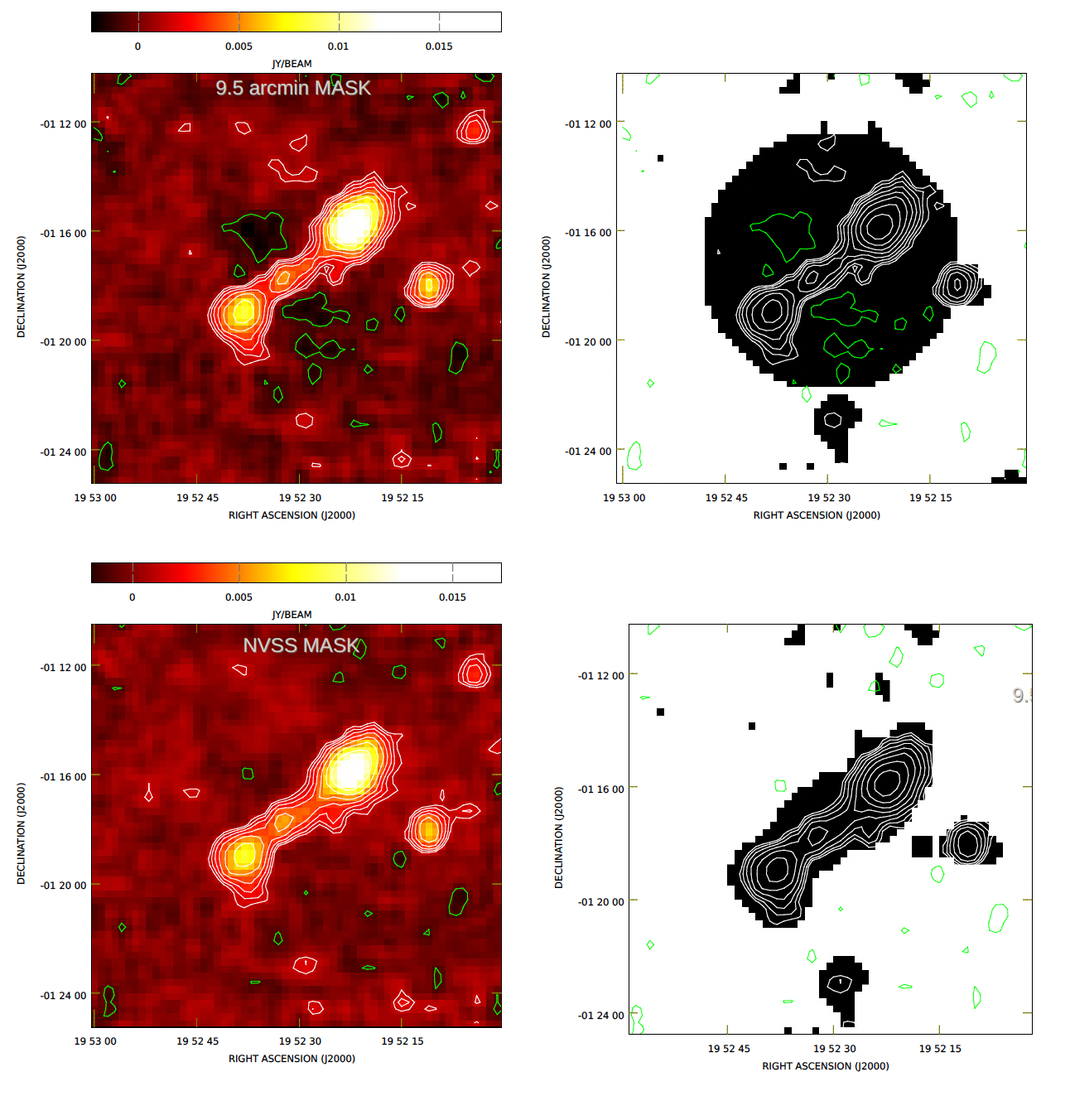}
		\caption{Images obtained using the circular mask with a diameter of 9.5\arcmin~ (\textit{top-left}) and the tailored NVSS mask (\textit{bottom-left}). The rms noise level is 0.39 mJy/beam for both images. Contours are traced at -3$\sigma$, 3$\sigma$ and scale by $\sqrt{2}$ (negative contours are represented in green). Top-right and bottom-right panels show the mask (black pixels) defining the cold-sky region (white pixels) used for the baseline fit.}
		\label{circular_nvss_mask}
	\end{figure*}

	We also repeated the baseline subtraction by using a smaller, more tailored, mask for 3CR\,403.1 at 18.6\,GHz based on the NVSS $3\sigma$ contour level, see Figure~\ref{circular_nvss_mask}. In this case the negative signal disappears. However, this test is not conclusive since the negative signal could have been captured and removed along with the cold sky emission. We concluded that new observations over a larger field of view are probably necessary	to firmly determine the nature of the negative radio brightness.

	\clearpage
	\bibliography{biblio}{}
	\bibliographystyle{aasjournal}
\end{document}